\documentclass[aps,prb,reprint,amssymb,amsmath]{revtex4-2}

\usepackage{xcolor}
\usepackage{soul} 
\usepackage{amsmath}
\usepackage[english]{babel} 
\usepackage{bm}
\usepackage{blindtext}
\usepackage{float}
\usepackage[T1]{fontenc}    
\usepackage[margin=1in,footskip=0.25in]{geometry}   
\usepackage{graphicx}       
\usepackage{multirow}       %
\graphicspath{{figures/}}
\DeclareMathOperator{\sech}{sech}
\newcommand*{\rttensor}[1]{\overline{\overline{#1}}}

\begin{document}

\preprint{APS/123-QED}

\title{Internal mechanical dissipation mechanisms in amorphous silicon}

\author{Carl L\'{e}vesque}
\email{carl.levesque@umontreal.ca}

\author{Sjoerd Roorda}%
\email{sjoerd.roorda@umontreal.ca}

\author{Fran\c{c}ois Schiettekatte}%
\email{francois.schiettekatte@umontreal.ca}

\author{Normand Mousseau}%
\email{normand.mousseau@umontreal.ca}

\affiliation{%
D\'{e}partement de Physique and Regroupement qu\'{e}b\'{e}cois sur les mat\'{e}riaux de pointe, Universit\'{e} de Montr\'{e}al, C.P. 6128, succursale centre-ville, Montr\'{e}al, Qu\'{e}bec, Canada H3C 3J7
}

\begin{abstract}

Using the Activation-Relaxation Technique-nouveau, we search for two-level systems (TLSs) in models of  amorphous silicon (\emph{a}-Si). The TLSs are mechanisms related to internal mechanical dissipation and represent the main source of noise in the most sensitive frequency range of the largest gravitational wave detectors as well as one of the main sources of decoherence in many quantum computers. We show that in \emph{a}-Si, the majority of the TLSs of interest fall into two main categories: bond-defect hopping where neighbors exchange a topological defect and the Wooten-Winer-Weaire bond exchange. The distribution of these categories depends heavily on the preparation schedule of the \emph{a}-Si. We use our results to compute the mechanical loss in amorphous silicon, leading to a loss angle of 10$^{-3}$ at room temperature, decreasing to 10$^{-4}$ at 150 K in some configurations.  Our modeling results indicate that multiple classes of events can cause experimentally-relevant TLSs in disordered materials and, therefore, multiple attenuation strategies might be needed to reduce their impact.

\end{abstract}

\maketitle

\section{Introduction}
Current gravitational wave detectors (GWD) consist of Michelson interferometers with arms the length of a few km containing a Fabry-Perot cavity. Since 2015 \cite{Abbott_2015}, GWDs have successfully detected nearly a hundred events\cite{GWTC3}, at an accelerating pace thanks to continuous efforts. Among the targets for improvement are the test masses: massive dielectric mirrors at the end of each arm. Their reflective surface consist of a stack of alternating high refractive (HR)  and low refractive (LR) index materials. In the current implementation of LIGO and VIRGO, the LR material for coating is amorphous silica and the HR material Ti-doped amorphous tantala \cite{harry_2006}. Amorphous materials, especially in the HR layers, present intrinsic  fluctuations that can be directly linked to the mechanical dissipation ($Q^{-1}$) and thermal noise through the fluctuation-dissipation theorem \cite{kubo_fluctuation-dissipation_1966,saulson_thermal_1990, Levin_1998}. Such phenomena also cause decoherence in some quantum computers~\cite{Palma1996}. Despite considerable efforts to reduce these losses in the HR layers \cite{Abernathy_2021,granata_progress_2020,Lalande_2021}, low mechanical loss remains the limiting factor of noise at frequencies around 50 Hz in major GWDs, at which these detectors are the most sensitive \cite{nawrodt_challenges_2011}. 

Here, we investigate the  origin of the mechanisms leading to internal mechanical dissipation through atomistic simulations. In order to simplify the problem, we investigate amorphous silicon (\emph{a}-Si), a prototypical model of a continuous random network. \emph{A}-Si consists of a single element and its structure and dynamics have been under investigation for more than 50 years \cite{Lewis_2022}. Beyond its generic interest, \emph{a}-Si is directly relevant in the GWD context as it is considered for future generations of GWD \cite{Birney_2018}: it features a high refractive index, reducing the number of layers in the stack, and it can be synthesized with a ultra-low internal mechanical dissipation \cite{liu_hydrogen-free_2014}.  

We show that (i) a distribution of two-level systems within the experimentally-relevant energy and frequency range; (ii) depending on the relaxation state of the material, relevant events are dominated by a dangling bond hop or the more elaborated Wooten-Winer-Weaire (WWW)~\cite{wooten_computer_1985} bond exchange mechanism; and (iii) the loss angle deduced from our models is compatible with experiments.  Overall, these results suggest that since multiple classes of events can be associated with two-level systems in the same frequency range in disordered materials, so multiple strategies might be needed to reduce their number.

\section{Theory and methods}
\subsection{Two-level systems}\label{TLS_theory}

Due to metastability associated with  structural disorder and a distribution in local strain, amorphous materials present, intrinsically, more possibilities for the presence of local minima separated by low-energy barriers than their crystalline counterparts. As they evolve over time, solid-state systems can be pictured as transiting from one local minimum to the other on a potential energy landscape~\cite{wales_energy_2004}.   In the context of amorphous solids, a two-level system happens when two minima are connected by a low-energy single saddle point and surrounded by much higher energy barriers so that the system is locally trapped in a two-state basin. 

\begin{figure}[tb]
    \centering
    \includegraphics[scale=0.4]{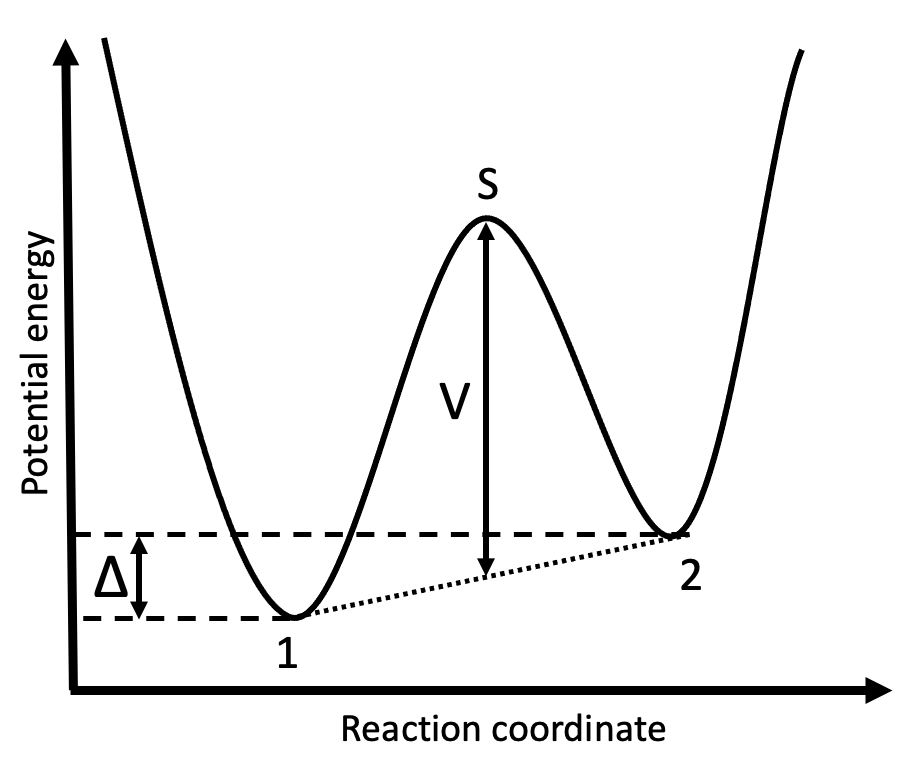}
    \caption{Potential energy landscape representation of a two-level system. The minima 1 and 2 are connected by a saddle point S. The mean barrier is V and the asymmetry is $\Delta$.}
    \label{TLS_rep}
\end{figure}

Two-level systems (TLSs) in amorphous solids were first introduced as a way to explain the behavior of their specific heat and thermal conductivity at low temperatures \cite{Anderson_1972}. Atoms tunneling through saddle points of TLSs provide more degrees of freedom to the system, increasing its heat capacity. 
As illustrated in Fig. \ref{TLS_rep}, a TLS can be characterized by the energy of its saddle point (S) relative to the minima (states 1 and 2), called the barrier ($V$), and the difference in energy between the two minima, called the asymmetry ($\Delta$).

We note that while the rugged nature of the energy landscape of amorphous solids gives a very large, quasi-continuous global distribution of barriers and asymmetries~\cite{Kallel2010}, events are localized in space and distributed throughout all the sample. Therefore they can be trapped by important gaps in the local barrier distributions, making dissipation a local phenomenon.

Assuming that quantum tunneling is negligible and that the system has the time to thermalize in a local state  after crossing a saddle point, the mean rate at which it will transition from minimum 1 to minimum 2, $\tau_{12}$ is given by the Arrhenius law 
\begin{equation}\label{Arrhenius}
    \tau_{12} = \tau_0 e^{\frac{E_S - E_1}{k_B T}},
\end{equation}
where $\tau_0^{-1}$ is the attempt frequency, $E_s$ and $E_1$ are the energy levels at the saddle point and the first minimum, respectively. While activated prefactor can vary considerably~\cite{gelin_enthalpy-entropy_2020},  calculations of the attempt frequencies in \emph{a}-Si with the harmonic transition state theory were found to be close to $10^{13}$ s$^{-1}$ with small fluctuations\cite{valiquette_energy_2003}. To shorten the simulations, we shall use a constant value of $\tau_0 = 10^{-13}$ s.

Because of the finite asymmetry $\Delta$, $\tau_{12}$ is different from $\tau_{21}$. For the whole TLS, a relaxation time, $\tau$ is defined as a function of temperature $T, V, \Delta,$ and $\tau_{0}$ \cite{jackle_elastic_1976}
\begin{equation}\label{relax_time}
    \tau = \tau_0 \sech\left( \frac{\Delta}{2 k_B T} \right) e^{\frac{V}{k_B T}}.
\end{equation}

Mechanical dissipation in the frequency regime of Hz to MHz in amorphous materials, in response to a strain wave of frequency $\omega$, is widely thought to originate from the excitation of TLSs \cite{phillips_two-level_1987}. Mechanical energy from the wave ($V \pm \Delta/2$) can push the TLS to its saddle point. The TLS will then relax to state 2, transforming the mechanical energy into thermal energy.  Dissipation from a given TLS will be maximized when its relaxation time (Eq. \ref{relax_time}) matches the inverse frequency of the dissipated excitation. An approximation for relevant barriers is given by 

\begin{equation}\label{relevant_barriers}
    V \sim k_B T \ln{\frac{1}{\omega \tau_0}}.
\end{equation}

In the context of gravitational wave
detectors, mirrors are kept at temperatures ranging from room temperature down to cryogenic temperatures of 124 K \cite{birney_amorphous_2018}. 
At these temperatures and frequencies, quantum tunneling will be completely negligible in comparison to thermal activation, such that Eq. \ref{Arrhenius} is valid. Taking $\omega = 50$ Hz we get relevant barriers of 0.28 eV and 0.67 eV for temperatures of 124 K and 300 K, respectively.

If the entire distribution of TLSs configuration is known, the inverse quality factor, $Q^{-1}$, also called the loss angle of the bulk material, can be computed as follows:

\begin{equation}\label{Qsum}
    Q^{-1}_{( \omega )} = \frac{1}{E} \sum_{i} \frac{\gamma_i^2}{k_B T} \frac{\omega \tau_i}{1+\omega^2 \tau_i^2} \sech^2\left( \frac{\Delta_i}{2 k_B T} \right). 
\end{equation}

\noindent A detailed derivation for this equation can be found in Ref.~\cite{phillips_two-level_1987}. The sum runs over every TLS in the system. $\tau_i$ and $\Delta_i$ is the relaxation time and asymmetry of the TLS $i$. $E$ is the elastic modulus and $\omega$ is the frequency of the applied strain. $\gamma_i$ is the strength of the coupling between the TLS $i$ and the strain called the deformation potential. This deformation potential can be obtained from the coupling tensor
\begin{equation}\label{gamma_tensor}
    \rttensor{\gamma} = \partial \Delta / \partial \rttensor{\epsilon} ,
\end{equation}
where $\rttensor{\epsilon}$ is the strain tensor.

Both $\gamma$ and $E$ depend on the nature of the strain. For example, to compute the attenuation of a longitudinal wave with Eq.~\ref{Qsum}, $E$ will be the longitudinal modulus and $\gamma$ will be the longitudinal deformation potential. A detailed derivation has been carried out by Damart and Rodney in Ref. \cite{damart_atomistic_2018}. 

\subsection{Atomic models \label{Atomic_model}}

To conduct this study, we consider quenched-melt and hyperuniform network models. 

200  quenched-melt systems of 1000 atoms of \emph{a}-Si are generated. All 200 models are prepared using the molecular dynamics (MD) simulation software LAMMPS \cite{thompson_lammps_2022} following the same melt-quench procedure: 1000 atoms are distributed randomly in a periodic box at a temperature of 3000 K, the system is then cooled (quenched) at a (relatively) slow rate of $10^{11}$ K/s, freezing in an amorphous configuration. This method has the advantage of moderate computing cost and melt quench methods with slow cooling rate have been shown to generate samples that are in good agreement with well-annealed experimental \emph{a}-Si \cite{ishimaru_generation_1997,deringer_realistic_2018}.

We compare these models with a nearly hyperuniform network of \emph{a}-Si developed by Hejna \emph{et al.} \cite{hejna_nearly_2013}. This system was built using a modified version of the WWW algorithm \cite{wooten_computer_1985} developped by Barkema and Mousseau \cite{barkema_high-quality_2000}.  The systems are then annealed for further relaxation and meticulously compared to experimental data.  

To correctly simulate the structural and vibrational properties of \emph{a}-Si, a modified version of  the Stillinger-Weber potential parameter set, developed by Vink \emph{et al.} \cite{vink_fitting_2001} was used and applied to the original formulation.

For our topological analysis we define a cutoff value for two atoms to be connected at the middle point between the first and the second peak of the radial distribution function our our configurations. This middle point lies around 3.05 \AA. This is the same cutoff definition and value as Ref.~\cite{barkema_high-quality_2000}.

\subsection{Activation-Relaxation Technique}

In this work we search for thermally activated events with characteristic times of the order of the millisecond to the second. Here we select to used the Activation-Relaxation Technique nouveau (ARTn) \cite{barkema_event-based_1996,malek_dynamics_2000}, a saddle point search method that is ideally suited for such tasks as it focuses on finding high barrier events (with high characteristic times) without having to compute every thermal atomic vibration.

This technique samples events in the potential energy landscape and finds their barriers and asymmetries. Characteristic times are then found using Eq~\ref{relax_time}. MD-based methods can also be used to identify saddle points in such systems \cite{damart_atomistic_2018,hamdan_molecular_2014}. Because the timescale on which such method operates (~ns), only low-energy barriers are efficiently identified. Those contribute to dissipation at low temperature or high frequency according to Eq.~\ref{relevant_barriers}. Hence, both methods are complementary.

\section{Results and discussion}
\subsection{Sampling}

\begin{figure}[tb]
    \centering
    \includegraphics[width=\linewidth]{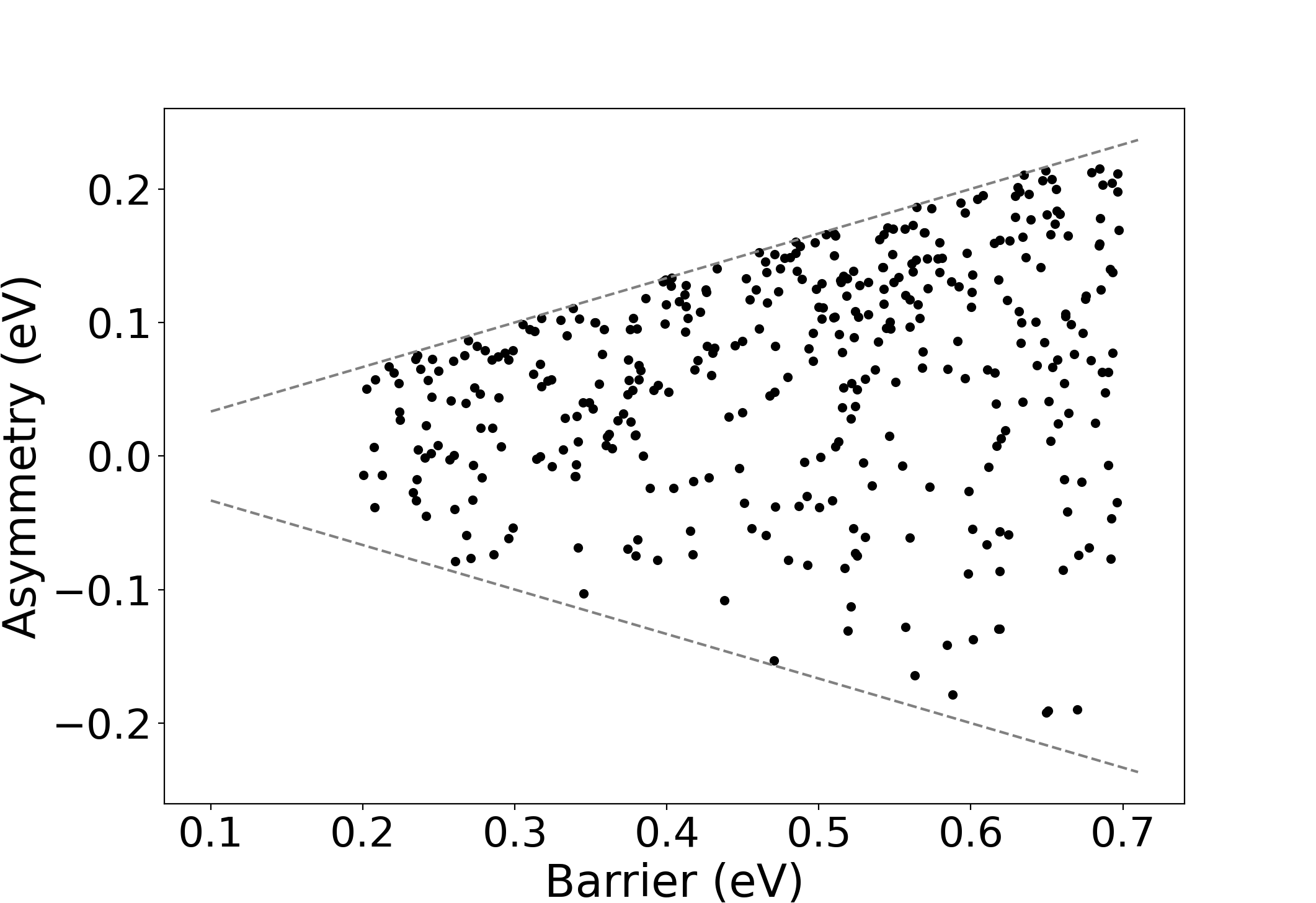}
    \caption{Barrier and asymmetry of events found by ART. The gray dotted lines show the asymmetry cutoff.}
    \label{TLS_dist}
\end{figure}

Using ARTn, we first explore the energy landscape around the final structure of the 200 quenched-melt 1000-atom configurations. For each sample, 30 ARTn searches are conducted per local topology. Since,  amorphous materials  the number of different environments is much larger than 1000, the number of local environments centered on the atoms in each of the systems, that means  30~000 event searches per configuration. While events with energy barriers ranging from 0 to a 5~eV cut-off are generated,  only events with an activated barrier $0.2 \leq V \leq 0.7$ eV are considered in the study because, as explained in section II-A, TLS with lower or higher barriers will not contribute significantly to internal mechanical dissipation  between temperatures of 124 and 300 K (Eq.~\ref{relevant_barriers}). A cutoff in asymmetry is chosen at $V = \Delta / 3$, represented by the diagonal lines in Fig.~\ref{TLS_dist}, because the contribution to dissipation of events with higher asymmetry is exponentially suppressed by large $\Delta$ (see Eq.~\ref{Qsum}). Damart and Rodney have also shown in \cite{damart_atomistic_2018} that TLSs with larger asymmetry do not contribute to the mechanical loss. The barrier and asymmetry of the remaining events are plotted in Fig.~\ref{TLS_dist}. We note that the TLS considered here represent only a very small fraction of all events found in this search.

TLS events found by ARTn can be characterized by the norm of the displacement of the main atom (i.e. the one moving the most) when the system transitions between two minima of the potential energy landscape, and the number of atoms involved in the transition. An atom is considered to be active if its displacement is greater than 0.1~\AA. 

\begin{figure}[tb]
    \centering
    \includegraphics[width=\linewidth]{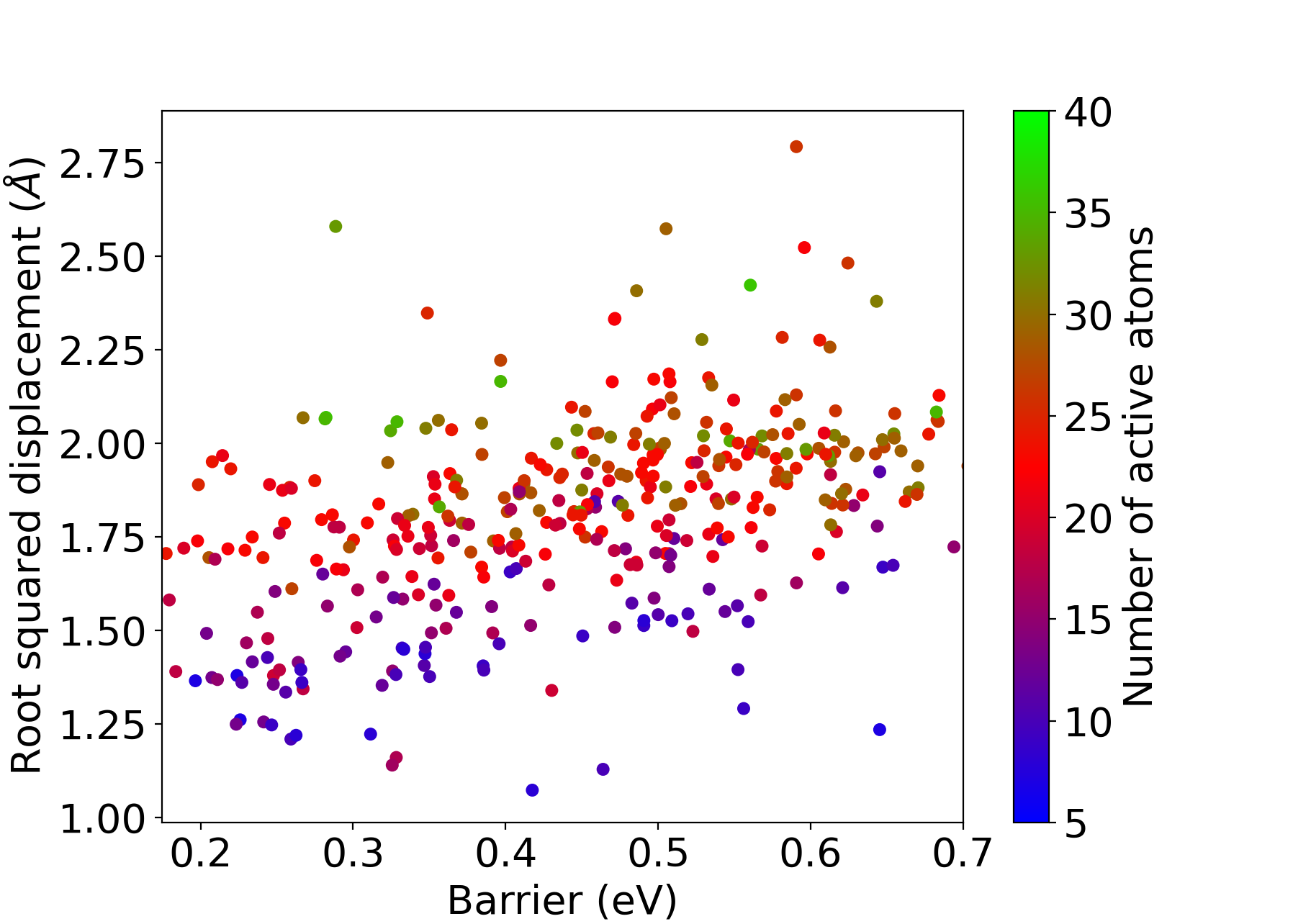}
    \caption{Root of squared atomic displacement in each TLS as a function of the energy barrier for the events plotted in Fig. \ref{TLS_dist}. Symbols are color-coded according to the number of active atoms during the event (right scale).}
    \label{displa}
\end{figure}

Fig.~\ref{displa} shows the root of the sum of squared displacements for active atoms; the marker color corresponds to the number of active atoms according to the scale on the right. We see that  the magnitude of the atomic displacements is correlated with the energy barrier, although with considerable dispersion. In addition,  large displacement for the main atoms is generally associated with  a larger number of active atoms (green dots). Conversely, the blue dots, representing events with a few active atoms, are at the bottom of the graph while the green dots, representing events with several active atoms, are at the top. 
In general, TLSs events involve between 5 to 30 active atoms, a number much smaller than the 20 to 150 active atoms  found in  in oxide glasses \cite{hamdan_molecular_2014, trinastic_molecular_2016}. This indicates that TLSs are much more localized phenomena in a more locally rigid structure with a higher coordination number, such as amorphous silicon, than oxides.

\subsection{Categorizing TLSs}

Following the classification of activated events in \emph{a}-Si developed by Barkema and Mousseau in \cite{barkema_identification_1998}, we adopt a three-class categorization based the evolution of the bond network during events.  The first category of TLSs (type 1) is associated with a bond hop from one atom to another. These events usually involve the diffusion of a coordination defect, such as dangling bond defects, and will be called \textit{bond defect hopping TLSs}. These jumps are typically made possible by the movement of a single atom, leading to a relaxation of the surrounding environment. 

\begin{figure}[tb]
    \centering
    \includegraphics[width=5cm]{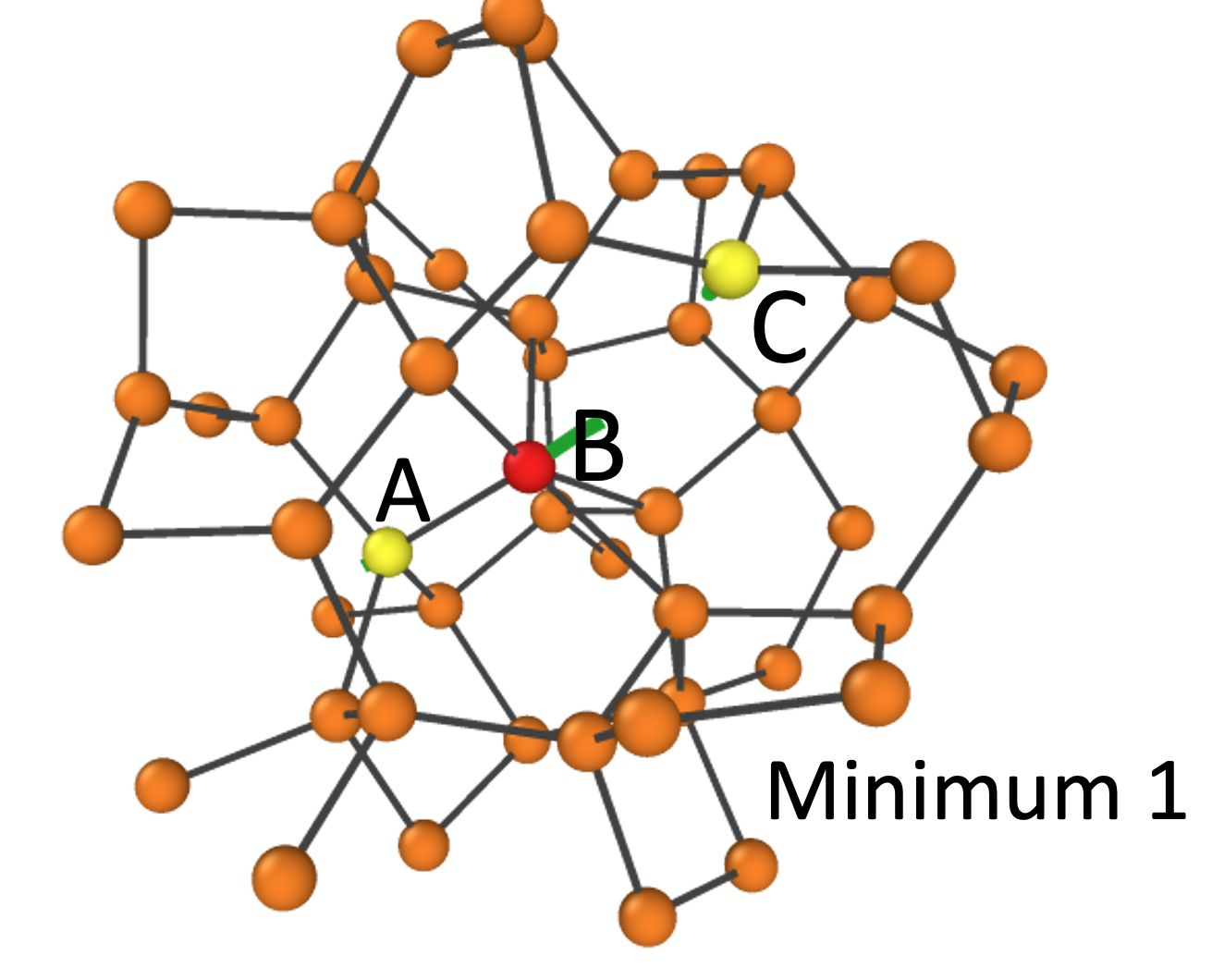}
    \includegraphics[width=5cm]{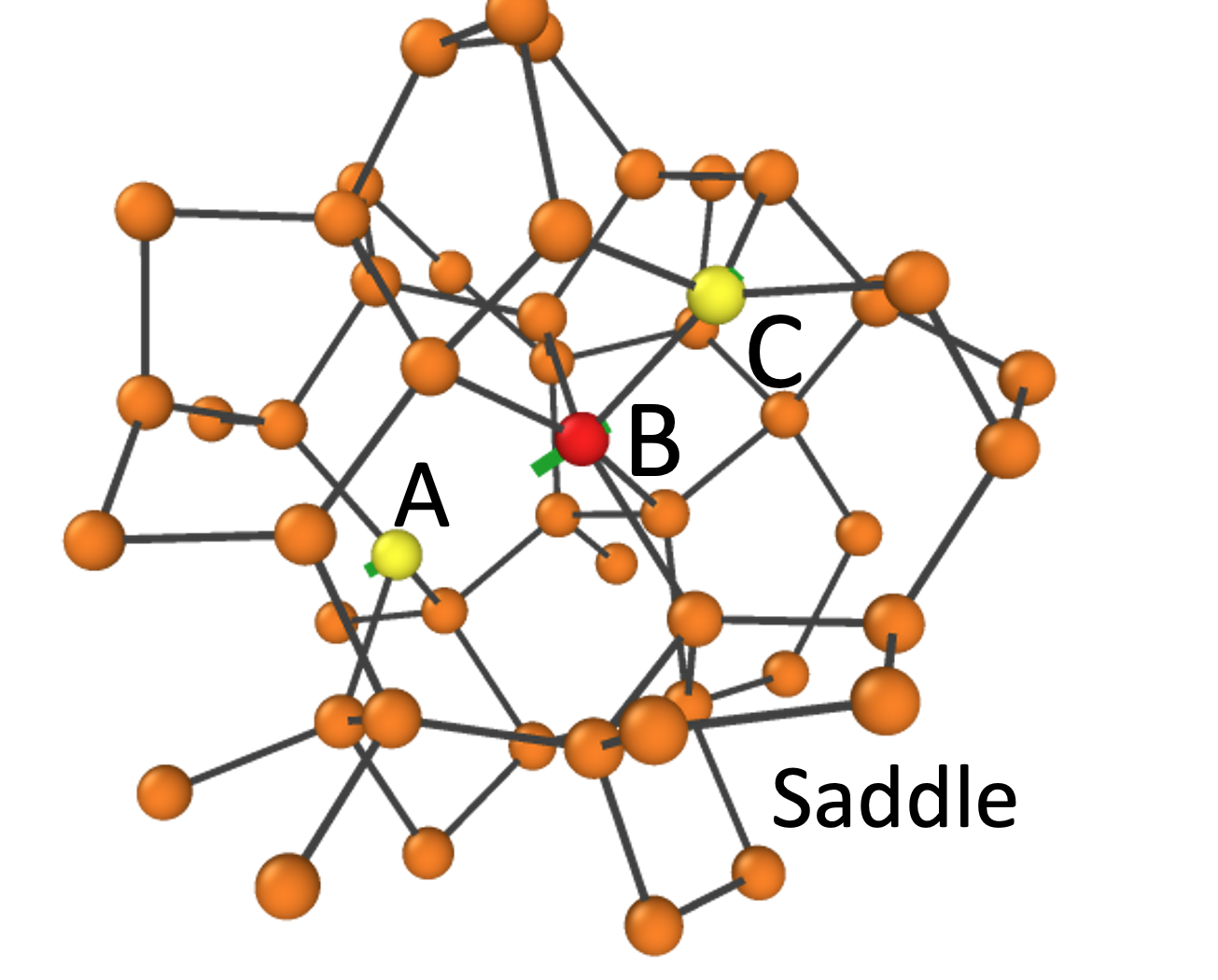}
    \includegraphics[width=5cm]{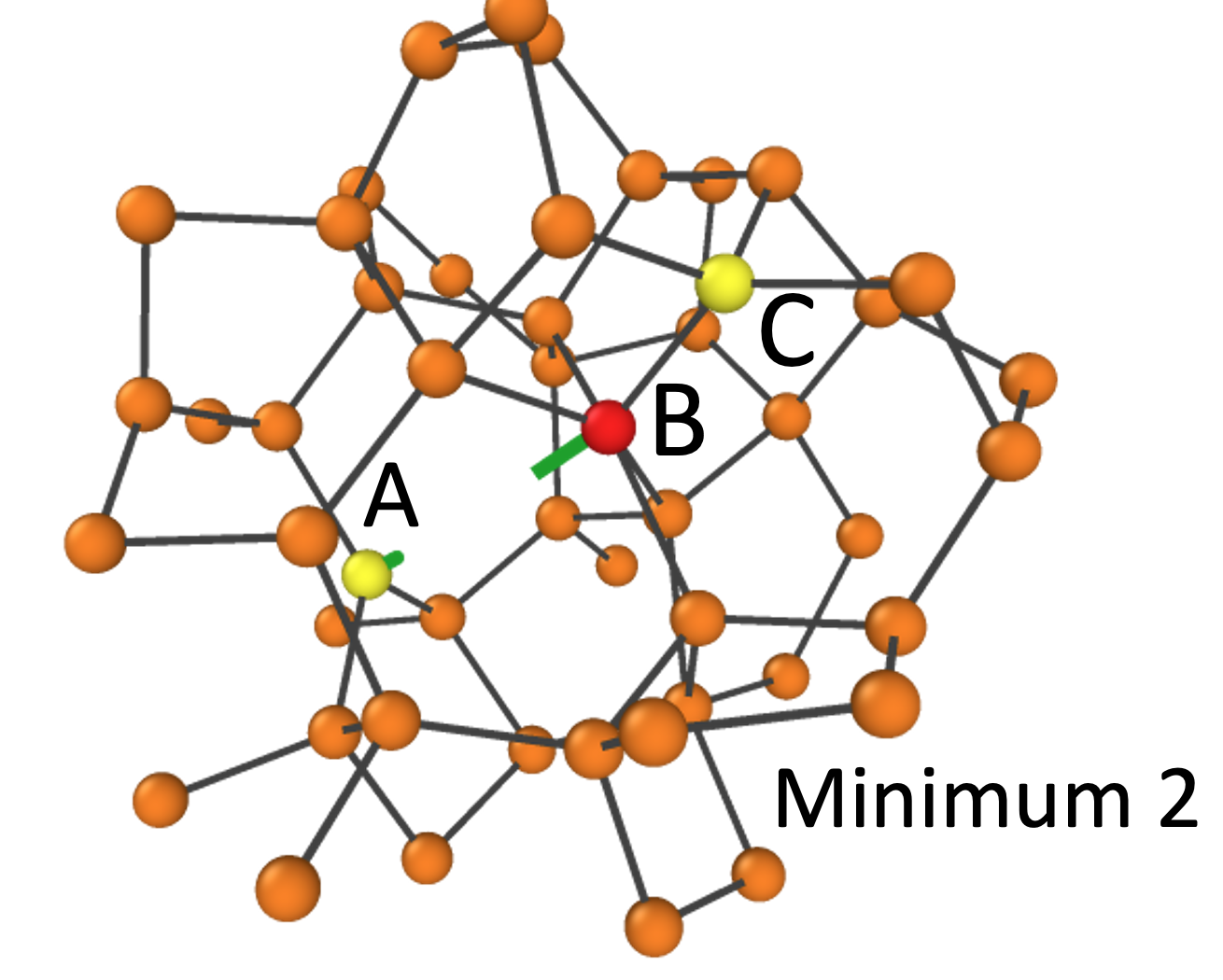}
    \caption{Example of bond defect hopping event (type 1). Panel a) and c) represent the initial and final minima, respectively, while panel  b) shows the system configuration at the saddle point. The central atom is in red. The atoms in yellow change their bonding status with the main atom. The green lines show the trajectory of the atoms between the frames.}
    \label{type1}
\end{figure}

In the example of Fig. \ref{type1}, the main atom (B, in red) is 5-fold coordinated. As for the yellow atoms, which are the ones that change their bond with the main atom during the process, atom A (in yellow) initially features coordination of 4, but becomes 3-fold coordinated at the end of the event. The inverse happens to atom C (also in yellow): it is initially 3-fold coordinated and becomes 4-fold coordinated during the process.   

\begin{figure}[tb]
    \centering
    \includegraphics[width=5cm]{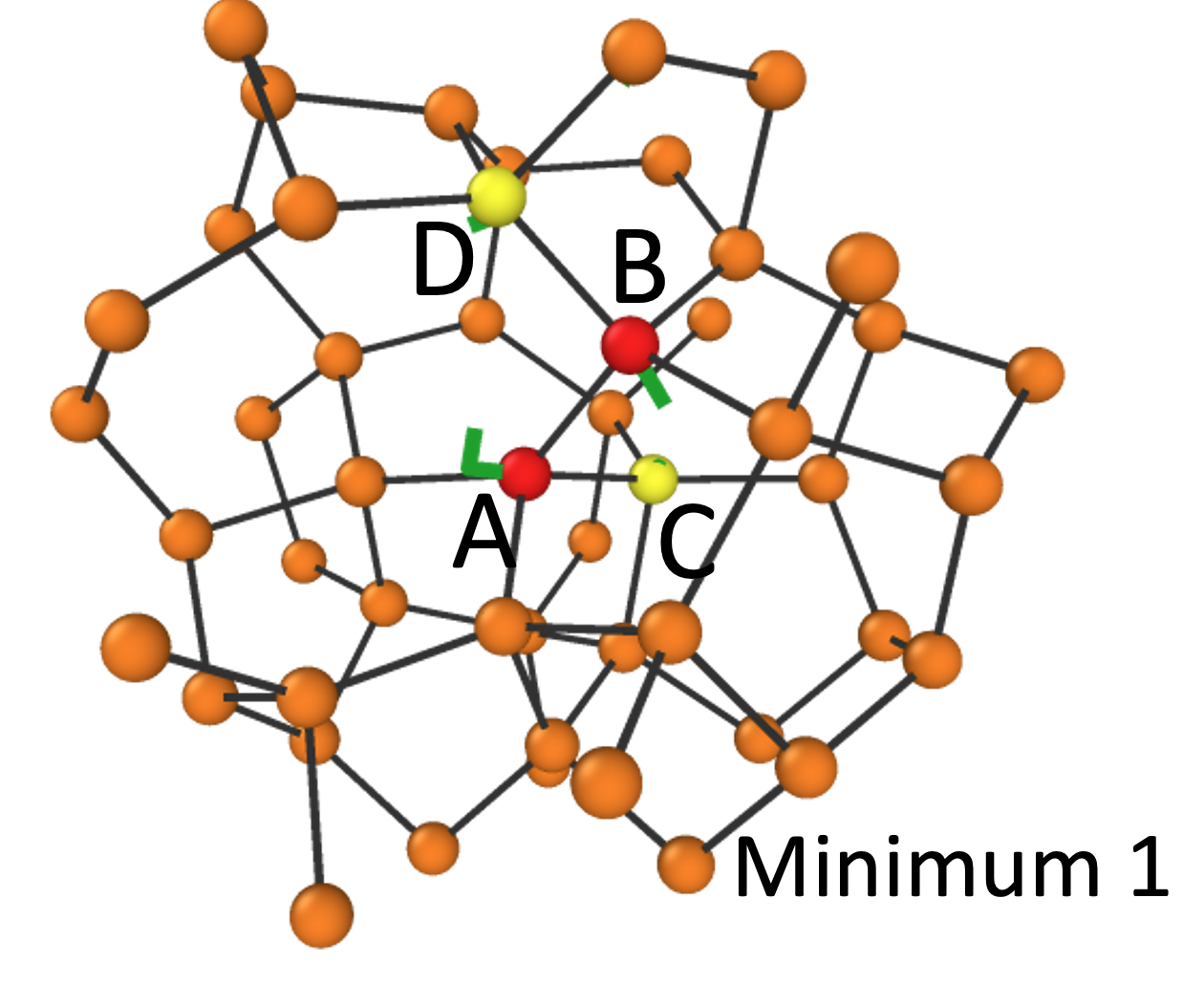}
    \includegraphics[width=5cm]{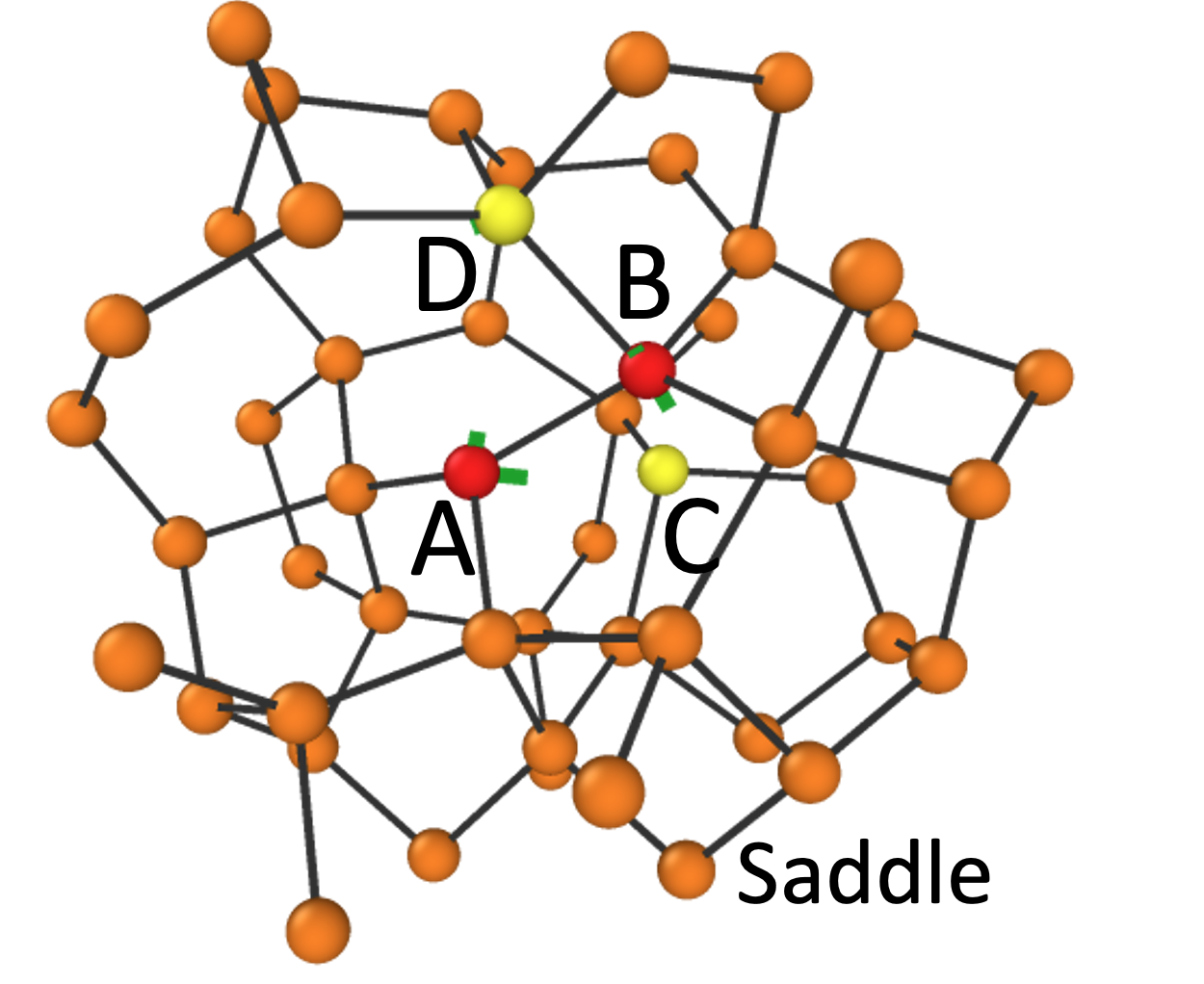}
    \includegraphics[width=5cm]{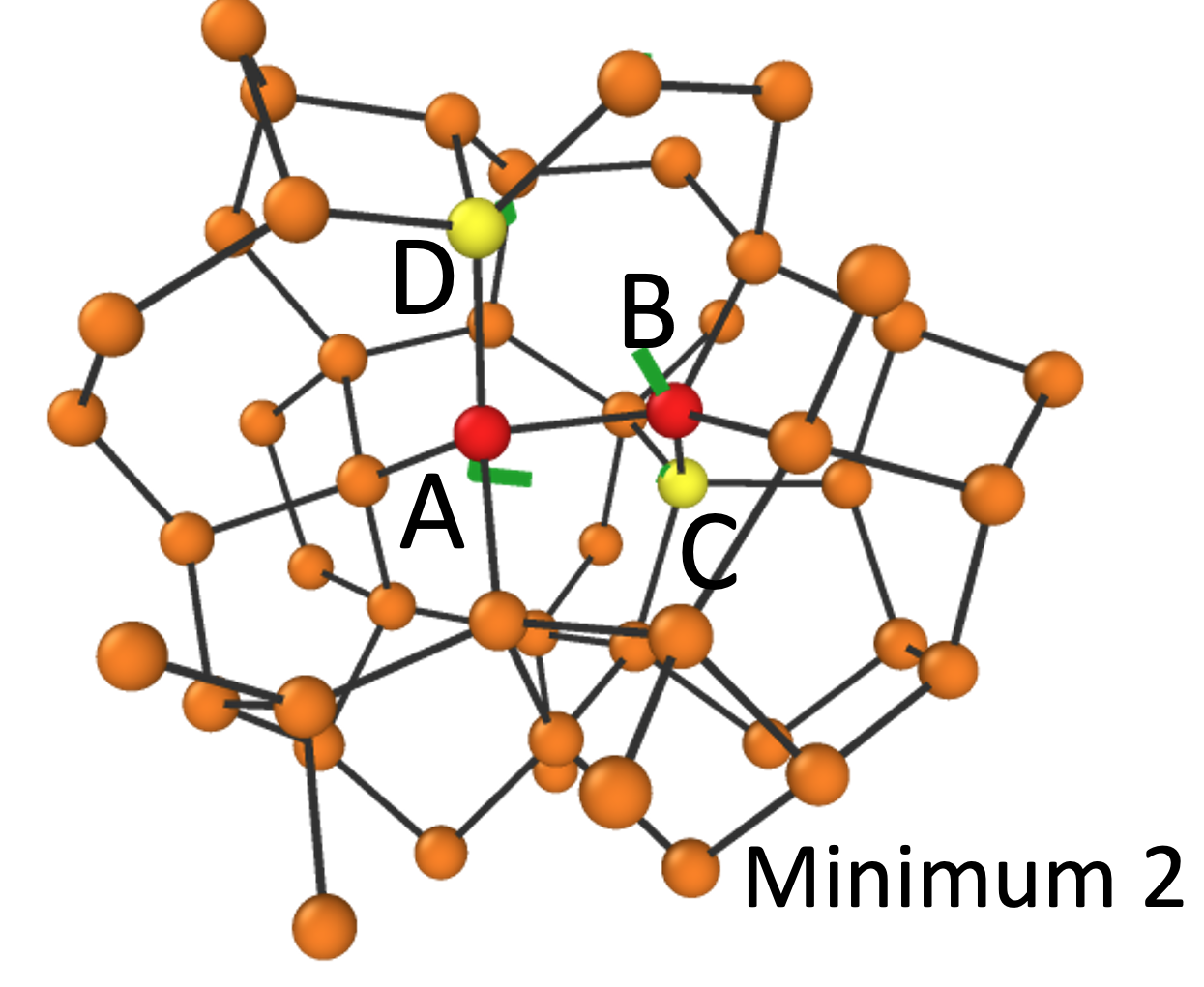}
    \caption{Example of a bond exchange TLS (type 2). The red atoms are the main atoms. The yellow atoms break a bond with one red atom and form a new bond with the other red atom. The green lines show the trajectory of the atoms between the frames.}
    \label{type2}
\end{figure}

The second category of TLSs (type 2) corresponds to the (WWW) bond exchange mechanism \cite{wooten_computer_1985} and is called \emph{bond-exchange TLS}.  This mechanism is commonly observed in \emph{a}-Si \cite{barkema_identification_1998,valiquette_energy_2003} and has been described in other amorphous and crystalline solids such as graphene\cite{stone1986}. It involves two connected atoms exchanging their respective bonds between them. An example is provided in Fig. \ref{type2}. The atoms A and B (in red) stay connected at all times and exchange their bonds to atoms C and D (in yellow). Interestingly, these events do not depend on the presence of a defect to occur, as opposed to type 1 events, and can occur in a material featuring few voids. 

The third class of events includes all TLSs that do not fit the first or the second category. Typically, these TLSs involve three or more atoms. While it is formally possible to further analyze them~\cite{mousseau2000,valiquette_energy_2003}, their diversity limits the understanding we can gain from their detailed classification so we will focus on type 1 and 2 TLSs.

\begin{figure}[tb]
    \centering
    \includegraphics[width=\linewidth]{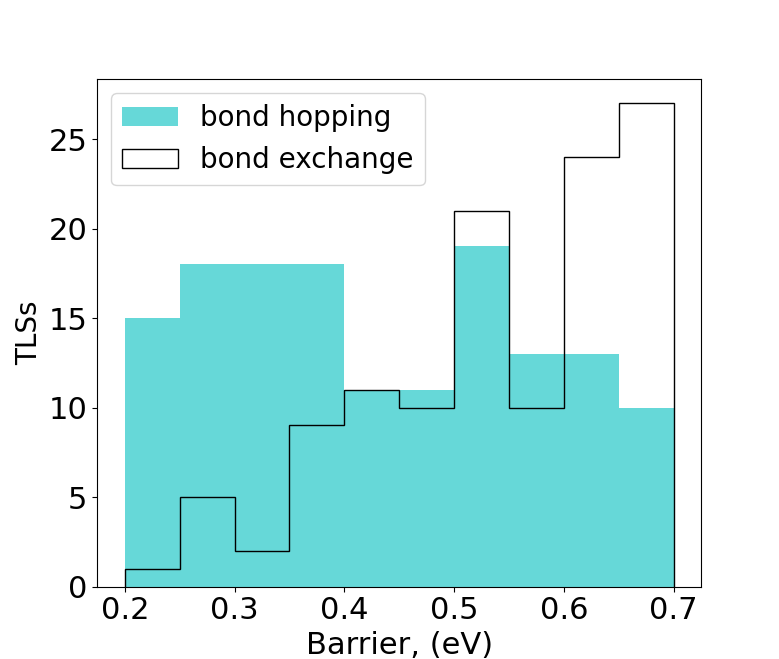}
    \caption{Barrier distribution of bond defect hopping (cyan, full bars) and bond exchange TLSs (empty bars).}
    \label{typeVbarr}
\end{figure}

Fig. \ref{typeVbarr} shows that bond defect hopping and bond exchange TLSs exhibit different barrier distributions. Bond defect hopping TLSs are associated with lower barriers than bond exchange, due to the presence of a bonding defect on the central atoms. The strain associated  these defects increases the potential energy of the two minima and lowers the saddle point barrier.  Asymmetries, on the other hand, are distributed relatively evenly for all types of TLSs. 

\subsection{Mean trends}

Local environments that support TLSs display some structural characteristics that separate them from the rest of the sample. 

\begin{figure}[tb]
    \centering
    \includegraphics[width=\linewidth]{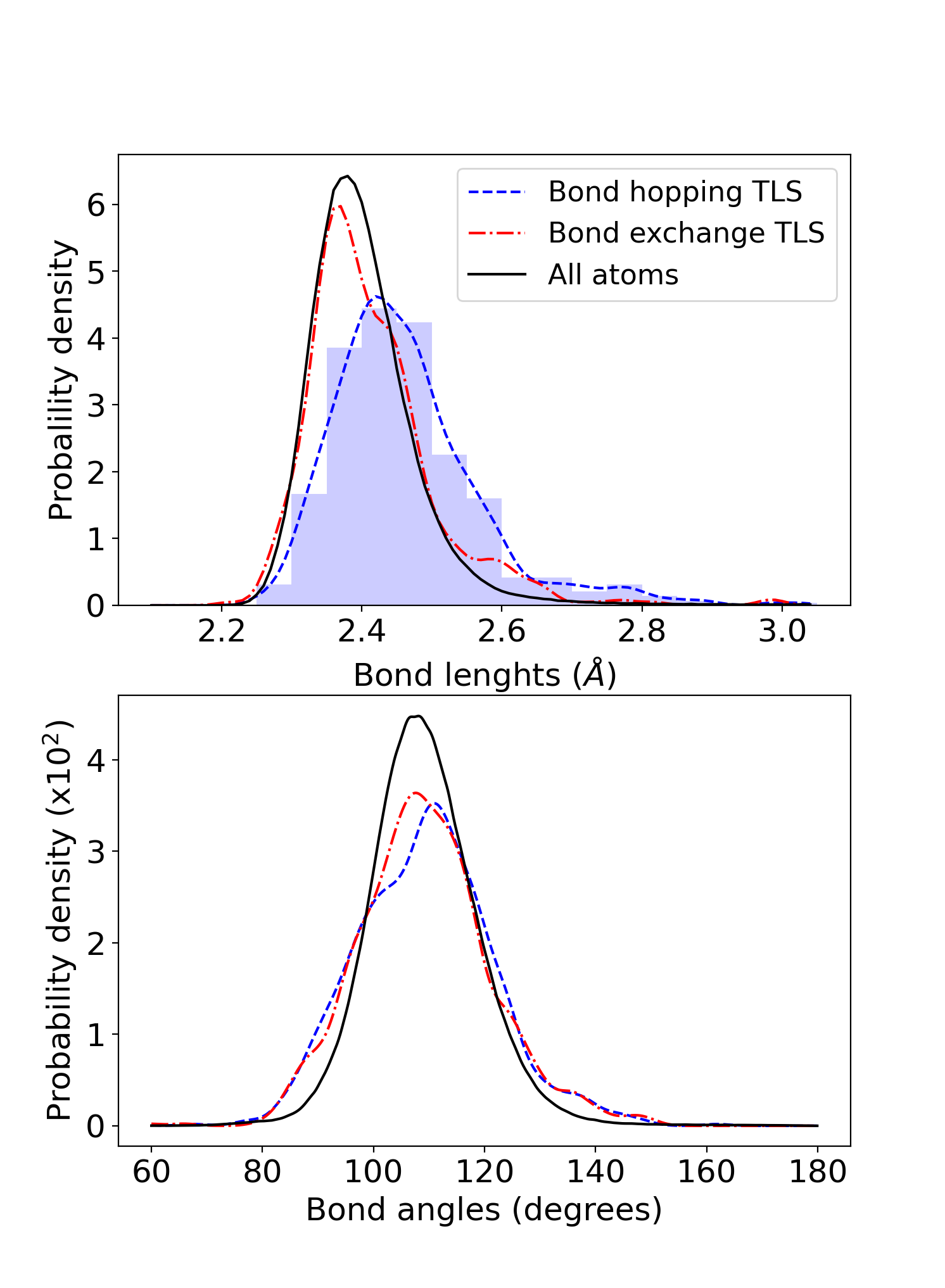}
    \caption{Smoothed distribution of the bond lengths (top) and the bond angles (bottom) of the main active atom in the bond defect hopping (blue dashed curves) and bond exchange TLSs (red dotted curves). The black dotted lines represent the first peak and bond angle distribution of the RDF of our \emph{a}-Si configurations. A raw histogram (blue, top) is also shown to illustrate the effect of the smoothing.}
    \label{lenghts_angles}
\end{figure}

These can be summarized by looking at the local density obtained from computing the Voronoi volume surrounding each atom. While the average density, as measured over all atoms, is 2.20~g/cm$^3$, the local value surrounding the dominant atom associated with TLSs is only 2.08~g/cm$^3$. These zones of low density are associated with strained or under-coordinated atoms, creating local instabilities that favor the formation of TLSs. Figure \ref{lenghts_angles} confirms this by showing that the dominant atom associated with TLSs have a longer bond (top graph, dashed curves) and a wider bond angle distributions (bottom graph, dashed curves) compared to all the atoms of the system (black solid curve).

\subsection{Relation between structure and mechanical loss}

Experiments have shown that thermal annealing or deposition at high substrate temperature reduces significantly the mechanical loss of amorphous materials \cite{Vajente_2018}, especially silicon \cite{liu_hydrogen-free_2014}. It is generally argued that this reduction is due to thermal activation that allows the material to reach more relaxed states. 

To assess the importance of this effect here, we  compare configurations obtained by different means. More specifically,  we compare the systems discussed until now, obtained by melt-quench models, with a nearly hyperuniform network (NHN) model built by Hejna \emph{et al.}  \cite{hejna_nearly_2013} and presented in section \ref{Atomic_model}. 

\begin{table}[tb]
\centering
\caption{Characteristics of \emph{a}-Si configurations prepared by melt-quench (this work) and by bond-exhange (NHN) \cite{hejna_nearly_2013, wooten_computer_1985}. Uncertainties in the first column correspond to the standard deviation computed over 200 independent samples. }
\begin{tabular}{c c c} 
    \hline
    Sample                    & Melt-Quench    & bond-exchange \\
     & & (NHN) \\
    \hline\hline
    Energy                    & \multirow{2}{*}{$-3.078 \pm 0.004$} & \multirow{2}{*}{-3.089} \\ 
    (eV per atom)             & & \\
    \hline
    Density                   & \multirow{2}{*}{$2.200 \pm 0.006$} & \multirow{2}{*}{2.21} \\ 
    (g/cm$^3$)                & & \\
    \hline
    Over-coordination defect  & \multirow{2}{*}{$17 \pm 4$}  & \multirow{2}{*}{0.4} \\ 
    (per 1000 atoms)           & & \\
    \hline
    Under-coordination defect & \multirow{2}{*}{$7 \pm 2$ }  & \multirow{2}{*}{0.45} \\ 
    (per 1000 atoms)           & & \\
    \hline
    TLSs found                 & & \\
    Nominal                    & 390  & 33   \\
    Per 1000 atoms             & 1.95 & 1.65 \\
    \hline\hline
\end{tabular}
\label{conf_compare}
\end{table}

We present some physical and structural characteristics of samples prepared using the two methods in Tab. \ref{conf_compare}. Despite similar atomic densities and potential energy per atom, the configuration obtained with a bond-exchange approach shows a significantly  lower density of point-defects compared to those obtained by melt-quench. 

\begin{figure}[tb]
    \centering
    \includegraphics[width=\linewidth]{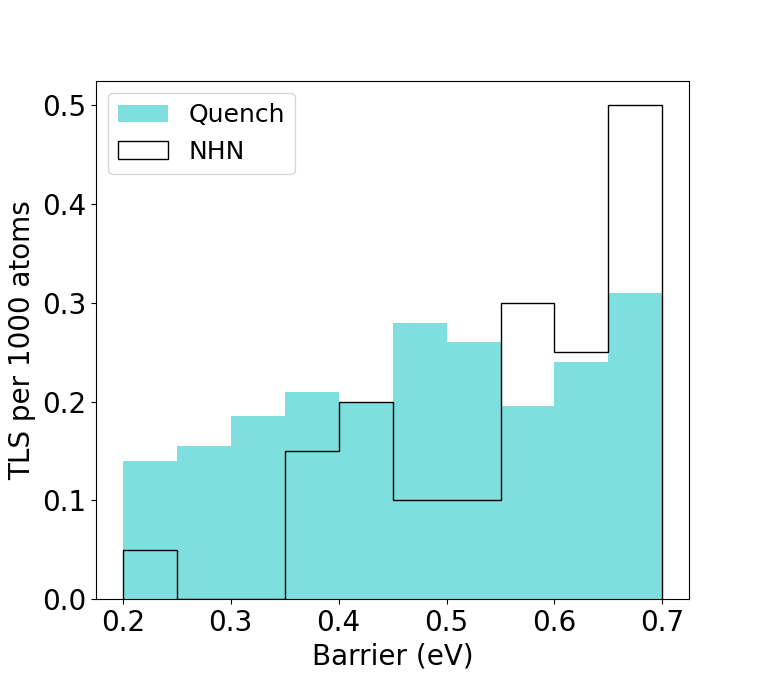}
    \caption{Energy barrier distribution for the TLSs in models obtained by the melt-quench and NHN preparation methods (described in text).}
    \label{TLS_relax}
\end{figure}

\begin{figure}[tb]
    \centering
    \includegraphics[width=\linewidth]{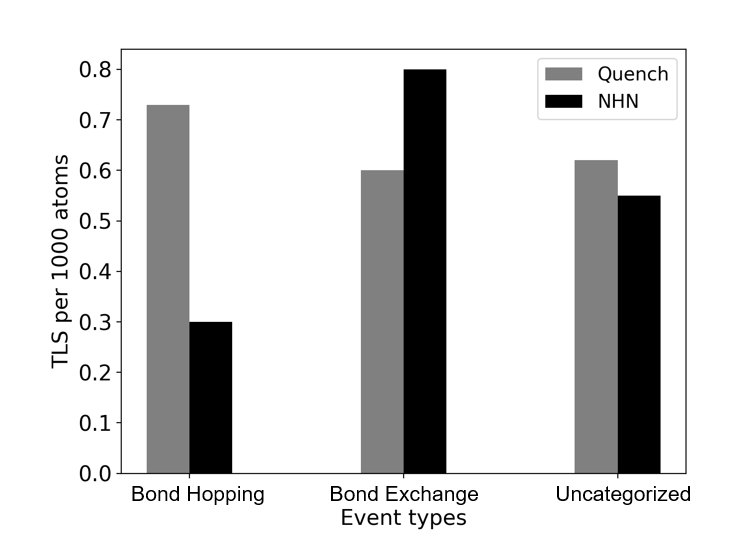}
    \caption{Density of the three types of TLSs in quenched \emph{a}-Si (blue) and NHN \emph{a}-Si (orange).}
    \label{types_hist}
\end{figure}

The lower density of the NHN configuration is reflected in its TLSs distribution generated with ARTn, as before. Figure \ref{TLS_relax} shows a depleted TLS distribution in the NHN configuration at low barriers (between 0.1 to 0.4 eV) as compared with the melt-quench models. This depletion is associated with a reduced ratio of bond defect hopping (type 1) to bond exchange (type 2) TLSs. This is not too surprising considering that bond hopping events have in general lower barriers (Fig. \ref{typeVbarr}) and often involve coordination defects, which are much rarer in the NHN model as seen from Tab.~\ref{conf_compare}. Figure \ref{types_hist} shows histograms of the number of events by type in the melt-quench and NHM models normalized by the number of atoms in each model. This confirms the much lower density of bond defect hopping event in NHN models, but it also underlines that the similar strain leads to a roughly similar total density of TLSs on both the NHN and the melt-quenched systems (also see Tab~\ref{conf_compare}).

\subsection{Loss angle calculations}

Knowing the microscopic details of TLSs, it is possible to compute directly the mechanical loss of \emph{a}-Si using Eq.~\ref{Qsum}. The details and approximations used are described in section \ref{TLS_theory}. 

The first step is to compute the strain-asymmetry coupling tensor in Eq~\ref{gamma_tensor}. We obtain this quantity  by applying a small affine deformation to both minima of every TLS. The potential energy of each minimum is then computed. This is done for both positive and negative strain and results are averaged over both. We verify that varying the amplitude of the deformation does not change the results. 

The deformation potentials for longitudinal and traverse strain waves  are then computed using the formula derived by Damart and Rodney \cite{damart_atomistic_2018}. For the longitudinal deformation potentials, we get values between 0.5 and 9.1 eV and a mean value of 4.1 eV and for the transverse case we get energies ranging from 0.4 to 7.8 eV with a mean value of 3.4 eV. Deformation potentials are found to be uncorrelated to the barrier or the asymmetry. Experiments have reported average transverse deformation potentials of 1 eV \cite{fefferman_elastic_2017}, however those were conducted below 1 K, so barriers involved were much lower than those studied in this work, so we suppose the minima of active TLSs to be closer in geometry and their deformation potential to be lower. 

\begin{figure}[tb]
    \centering
    \includegraphics[width=\linewidth]{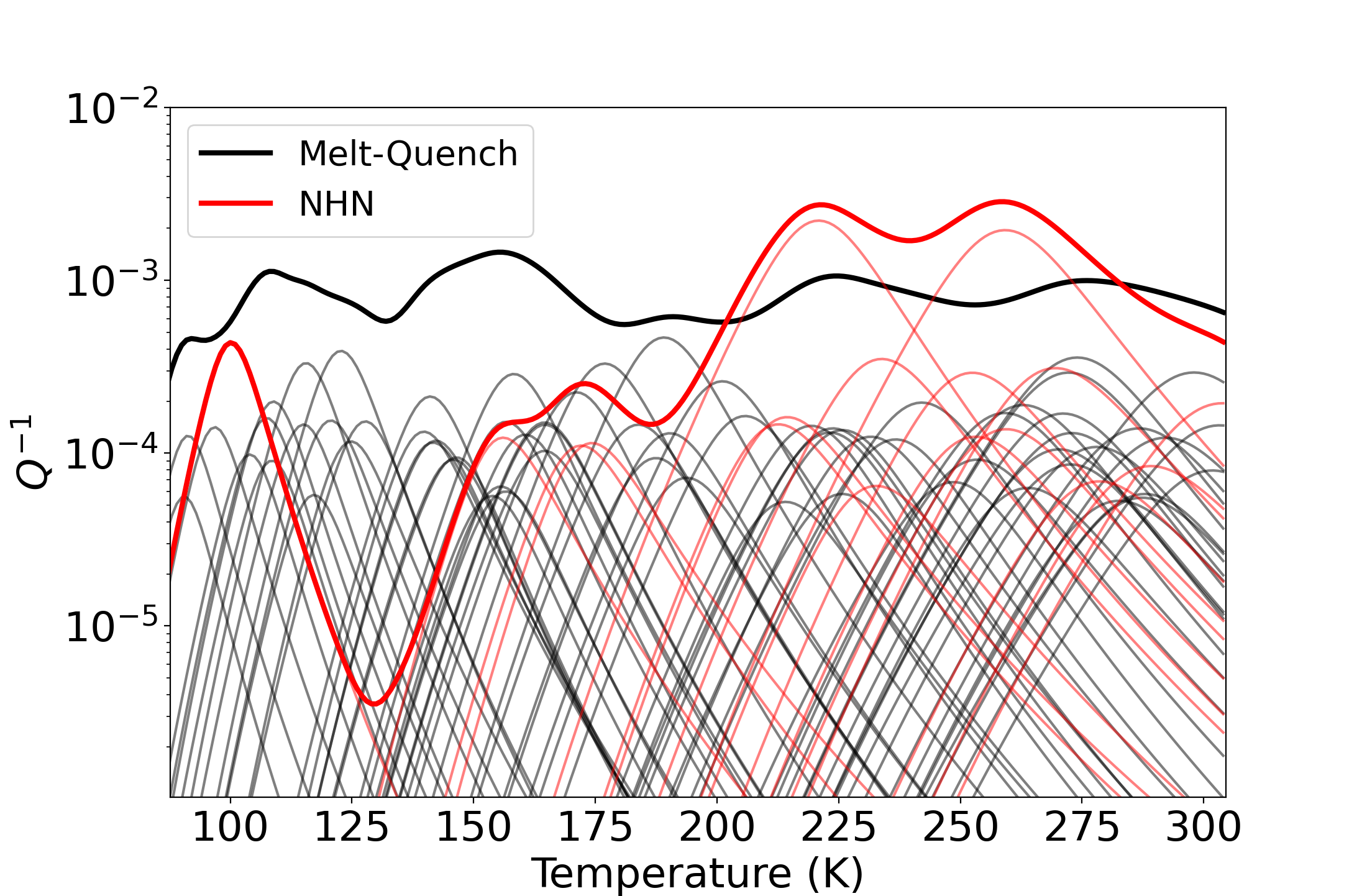}
    \caption{Internal mechanical dissipation computed for quenched \emph{a}-Si (black) and NHN \emph{a}-Si (red). The pale curves show the contributions from individual TLSs.}
    \label{IMD}
\end{figure}

Mechanical loss predictions are presented in Fig.~\ref{IMD} for both the melt-quench (black) and NHN configurations (red). The thin curves represent each term in the sum (Eq.~\ref{Qsum}), while the thick curve is the total. The melt-quench configurations lead to an almost constant $Q^{-1}$ value close to $10^{-3}$ at temperatures between 100 and 300 K. 

Because we consider only one sample, the NHN system presents a much lower absolute number of TLSs --- 33 vs 390, leading to more noise:   a few TLSs are causing peaks in the dissipation for this system, namely the ones centered at 100, 225 and 260 K. These TLSs have such high contributions  because of their very low asymmetry (0.009, 0.021 and 0.034 eV, respectively). We expect that these peaks would flatten with more data obtained on a larger system or better statistics. 

Nevertheless, it is remarkable that the NHN configuration yields a similar mechanical loss to that of the melt-quench around room temperature. Dropping down to cryogenic temperatures however, the mechanical loss decreases significantly, getting as low as $10^{-5}$ (although this result is strongly influence by single TLSs, such as the one producing a peak near 100 K). In the previous section we showed that the bond defect hopping TLSs, which compose most of the low barrier TLSs in \emph{a}-Si (Fig.~\ref{types_hist}), are much less frequent in the NHN configuration, causing the rarefaction of low barriers in this configuration (Fig~\ref{TLS_relax}.). The same explanation applies here as well, as low barriers are active at these low temperatures. Despite the smaller sample size of the NHN causing sharp peaks in the mechanical loss calculations, the reduction of low barrier TLSs is very clear (Table.~\ref{conf_compare}), therefore this decrease of the loss angle with temperature should be significant.

This behavior of $Q^{-1}$ in the NHN \emph{a}-Si is similar to that of experimental hydrogenated \emph{a}-Si, where the mechanical loss is high at room temperature but decreases significantly between 300 and 10K \cite{liu_amorphous_1997}. 

\section{Discussion and conclusion}

This study aims to identify  the structural origin of internal mechanical dissipation in amorphous solids. To do so, we focus on \emph{a}-Si, a classical reference for covalent disordered materials. More precisely, we characterize the two-level state mechanisms (TLSs) found in 200 sets of 1000 aotms of melt-quench generated configurations and  a single 20 000 atoms nearly hyperuniform
network (NHN) built built by  Hejna, Steinhardt and Torquato ~\cite{hejna_nearly_2013} using the WWW algorithm. Both configurations feature similar energies per atoms, however melt-quench generated \emph{a}-Si has around 2.5\% topological defects (3 and 5 fold cooridinated atoms), while the NHN shows almost none. The potentiel-energy landscape is explored with ARTn \cite{barkema_event-based_1996, malek_dynamics_2000}.
We keep only TLSs with small barriers (0.2-0.7 eV) and asymmetries (< $V/3$) because they correspond to the experimental observation window. 390 TLSs in the melt-quench \emph{a}-Si and 33 in the NHN are considered relevant to our study and further analyzed. 

Loss angle calculations on these systems lead to a high loss angle close to $10^{-3}$ for both NHN and melt-quench \emph{a}-Si at room temperature. While the loss angle in the melt-quench configurations stays relatively constant with temperature, NHN \emph{a}-Si shows an important decrease in loss angle when the temperature drops from 300 to 100 K, a similar behavior to that observed experimentally in well-relaxed \emph{a}-Si~\cite{liu_hydrogen-free_2014}. 

With the detailed information obtained through ARTn,  TLSs can be classified by the bond-network change associated with the two-level system. Two-thirds of events fall into two categories : bond defect hooping and bond exchange. These two types of TLSs are associated with  different local configurations: bond defect hopping happen around coordination defects and where bonds are stretched; and bond exchange TLSs are found in regions where all atoms are 4-fold coordinated but present small angle defects.

Our simulations demonstrate that various classes of TLSs can occur with different concentrations according to the preparation schedule.  Quenching from a melt result in bond defect hopping being the dominant TLS type, because of the high concentration of trapped point defects in these configurations. NHN of \emph{a}-Si has next to no point defects. This drastically reduces the concentration of bond hopping TLSs but has has little effect on the concentration of bond exchange TLSs. From this we learn that TLSs are independent: the presence of one type of TLSs do not depend on the existence of other types. We also show that types of TLSs we find in different configurations are also specific to the preparation, or the relaxation path of said configuration. 

Investigating other systems is necessary to improve our understanding of the atomistic origin of TLSs and internal mechanical dissipation. Work on oxide glasses was done in Ref. \cite{hamdan_molecular_2014, trinastic_molecular_2016, damart_atomistic_2018}. In Ref. \cite{damart_atomistic_2018}, Damart and Rodney analyzed TLSs in SiO$_2$, and found different archetypes of TLSs than in \emph{a}-Si, such as rotations of Si-O-Si chains. This suggests that  TLSs are  also specific to the the nature of the material, for instance between a-Si and and oxide glasses such as SiO$_2$ and Ta$_2$O$_5$. 

Decreasing TLSs in amorphous materials, means therefore understanding the nature of events for specific systems, but also addressing separately each of the potential classes of TLSs. 

Because of the importance of specificity, further study would benefit from better physical description to ensure that the atomistic details of TLSs are accurate. Since  larger amorphous configurations tend to better reproduce experimental observations \cite{barkema_high-quality_2000, igram_large_2018}, \textit{ab initio} approaches are probably not appropriate. Recent development in machine-learning forcefields~\cite{deringer_realistic_2018} offer here, new opportunity to deepen our understanding of these fascinating problems.

\section{Code and data availability}

The ARTn packages as well as the data reported here are distributed freely. Please contact Normand Mousseau (normand.mousseau@umontreal.ca).

\begin{acknowledgments}

The authors would like to thank K. Prasai and the other members of the Optical Working Group of the LIGO Scientific Collaboration for fruitful discussions. They also thank M. Hejna \emph{et al.} for sharing their NHN model. This project is supported by a team grant from the Fonds de recherche du Québec - Nature et technologie. NM and FS acknowledges partial support through Discovery grants from the Natural Science and Engineering Research Council of Canada (NSERC). CL holds a NSERC graduate scholarship. We are grateful to Calcul Québec and Compute Canada for generous allocation of computational resources. 

\end{acknowledgments}

\bibliography{refs} 
\bibliographystyle{apsrev4-2}

\end{document}